
\input harvmac
\input epsf

\newcount\figno
\figno=0
\def\fig#1#2#3{
\par\begingroup\parindent=0pt\leftskip=1cm\rightskip=1cm\parindent=0pt
\baselineskip=11pt
\global\advance\figno by 1
\midinsert
\epsfxsize=#3
\centerline{\epsfbox{#2}}
\vskip 12pt
{\bf Figure \the\figno:} #1\par
\endinsert\endgroup\par
}
\def\figlabel#1{\xdef#1{\the\figno}}
\def\encadremath#1{\vbox{\hrule\hbox{\vrule\kern8pt\vbox{\kern8pt
\hbox{$\displaystyle #1$}\kern8pt}
\kern8pt\vrule}\hrule}}

\batchmode
  \font\bbbfont=msbm10
\errorstopmode
\newif\ifamsf\amsftrue
\ifx\bbbfont\nullfont
  \amsffalse
\fi
\ifamsf
\def\IR{\hbox{\bbbfont R}}
\def\IZ{\hbox{\bbbfont Z}}
\def\IF{\hbox{\bbbfont F}}
\def\IP{\hbox{\bbbfont P}}
\else
\def\IR{\relax{\rm I\kern-.18em R}}
\def\IZ{\relax\ifmmode\hbox{Z\kern-.4em Z}\else{Z\kern-.4em Z}\fi}
\def\IF{\relax{\rm I\kern-.18em F}}
\def\IP{\relax{\rm I\kern-.18em P}}
\fi

\overfullrule=0pt
\def\Title#1#2{\rightline{#1}\ifx\answ\bigans\nopagenumbers\pageno0\vskip1in
\else\pageno1\vskip.8in\fi \centerline{\titlefont #2}\vskip .5in}

%

\def\p{\partial}
%
%
\lref\thooft{G. 't Hooft, {\it A planar diagram theory for strong 
interactions}, Nucl. Phys. {\bf B72} (1974) 461.}
\lref\polyakov{ A. Polyakov, {\it String theory and quark confinement},
hep-th/9711002.}
\lref\mal{J. Maldacena, {\it The large N limit of superconformal
field theories and supergravity}, hep-th/9711200.}

\lref\wilson{ K. Wilson, Phys. Rep. {\bf 23} (1975) 331.}

\lref\camabi{C. Callan and J. Maldacena, {\it Brane dynamics from the
Born Infeld action}, C. Callan and J. Maldacena. 
hep-th/9708147.  }

\lref\ori{ O. Ganor,
{\it Six dimensional tensionless strings in the large N limit},
 Nucl. Phys. {\bf B489} (1997) 95,
 hep-th/9605201.}
\lref\imsy{ N. Itzhaki, J. Maldacena, J. Sonnenschein and S. 
Yankielowicz, {\it Supergravity and the large N limit of 
theories with 16 supercharges}, hep-th/9802042. }
\lref\wittenhol{E. Witten, {\it Anti de Sitter space and holography}, 
hep-th/9802150.}
\lref\gkp{ S. Gubser, I. Klebanov and A. Polyakov,
{\it Gauge theory correlators from noncritical string theory},
 hep-th/9802109.}


\lref\dthree{
I. Klebanov, {\it Worldvolume approach to absorption by nondilatonic 
branes}, Nucl. Phys. {\bf B499} (1997) 217;
S. Gubser, I. Klebanov and A. Tseytlin,
{\it String theory and classical absorption by three-branes},
Nucl. Phys. {\bf B499} (1997) 217,
hep-th/9703040;
 J. Maldacena and A. Strominger, 
{\it Universal low energy dynamics for rotating black holes},   
Phys. Rev. {\bf D56} (1997) 4975,
hep-th/9702015; 
S. Gubser and I. Klebanov, {\it Absorption by branes and Schwinger terms
in the world volume theory}, Phys. Lett. {\bf B413} (1997) 41. 
}

\lref\skes{
S. Kachru and E. Silverstein, {\it 4d conformal field theories and
strings on orbifolds}, hep-th/9802183.}

\lref\zerotwo{E. Witten, Proceedings of Strings 95, hep-th/9507121; 
 A. Strominger, {\it Open p-branes}, 
  Phys. Lett. B {\bf 383} (1996) 44,
hep-th/9512059;
N. Seiberg, {\it Non-trivial fixed points of the renormalization
group in six dimensions}, 
Phys. Lett. B {\bf 390} (1996) 169, hep-th/9609161;
N. Seiberg and E. Witten, {\it Comments on string dynamics in 
six dimensions},  Nucl. Phys. B{\bf 471} (1996) 121,
 hep-th/9603003.}

\lref\gradsh{I.S.~Gradshteyn and I.M.~Ryzhik, {\it Table of Integrals, 
Series, and Products}, Fifth Edition, A.~Jeffrey, ed. (Academic Press:
San Diego, 1994).}

\lref\ho{ G. Horowitz and H. Ooguri, {\it Spectrum of large N 
gauge theory from supergravity}, hep-th/9802116.}

\lref\steve{ T. Banks, S. Gubser and A. Peet.}

%
\Title{\vbox{\baselineskip12pt
\hbox{hep-th/9803002}\hbox{HUTP-98/A014 }}}
{\vbox{
{\centerline { Wilson loops in large $N$ field theories}}
  }}         
\centerline{Juan Maldacena\foot{malda@pauli.harvard.edu}}
\vskip.1in
\centerline{\it Lyman Laboratory of Physics, Harvard University,
Cambridge, MA 02138, USA}
\vskip.1in
\vskip .5in

\centerline{\bf Abstract}

We propose a method to calculate the expectation values of
an operator similar to the Wilson loop  in the large $N$ limit
of field theories. 
We consider ${\cal N} =4 $ 3+1 dimensional super-Yang-Mills.
The prescription involves 
calculating the area of a fundamental string worldsheet in 
certain supergravity backgrounds. 
We also consider the case of coincident M-theory fivebranes
where one is lead to calculating the area of M-theory 
two-branes. We briefly discuss the computation for 2+1 dimensional
super-Yang-Mills with sixteen supercharges which is non-conformal.
In all these cases we calculate the energy of quark-antiquark 
pair.

 \Date{}


\newsec{ Introduction}

It has been expected  for some time that the 't Hooft limit \thooft\ of large
$N$ gauge theories is related to a string theory (see {\polyakov}
 and references therein).
In \mal\ a precise string theory was proposed for the 't Hooft limit
of ${\cal N } = 4 $ super-Yang-Mills in 3+1 dimensions, based 
on earlier studies \refs{\dthree}.
The 't Hooft limit is defined as the limit of $N \to \infty$ keeping
$g^2_{YM}N$ fixed. 
In this limit we get weakly coupled string theory on $AdS_5\times S^5$
where the radius of the five-sphere and the curvature radius of 
anti-de Sitter are proportional to 
$(g^2_{YM}N)^{1/4}$ in string units. 
There is also a flux of the Ramond-Ramond self dual five-form
field strength on the five-sphere. 
The string coupling is $g \sim g^2_{YM}$ and goes to zero in the
t' Hooft limit. In general we do not know how to solve 
free string theory on $AdS_5\times S^5$. However, when the
$gN$ is large the radius of curvature is large and we can
use the string in background fields approximation. 
In \refs{\gkp,\ho,\wittenhol}
it was shown
 how to calculate conformal dimensions  of operators and 
correlators in conformal field theory in 
terms of supergravity when   $gN$ is large. 
In this paper we consider the problem of calculating 
the expectation values of Wilson loop operators. 
The proposal is that these expectation values 
correspond to the area of a worldsheet whose boundary 
is the loop in question. We will further consider
similar observables for the M5-brane theory (the conformal (0,2)
six dimensional theory). We  also discuss Wilson loops in
non-conformal theories associated with D-twobranes.

\newsec{The Wilson loop}

Consider a Yang-Mills theory. The Wilson loop operator is
\eqn\wilsonloop{
W({\cal C} ) = { 1 \over N} Tr P e^{i  \oint_{\cal C} A} 
}
where ${\cal C}$ denotes a closed loop in spacetime and the 
trace is over the fundamental representation. 
We will be considering mostly the Euclidean field theory.
We can view the Wilson loop as the phase factor associated
to the propagation of 
a very massive quark in the fundamental representation
of the gauge group. 
A loop that is often considered is a rectangle as indicated
in figure 1 .  From 
the expectation value of this rectangular  Wilson loop
it is possible to read off the energy of a quark-antiquark 
pair. Namely, in the limit  $T \to \infty$ the expectation value of the 
Wilson loop is 
\eqn\larget{
\langle W({\cal C} ) \rangle  = A(L)  e^{ - T E(L) }
}
where $E(L)$ is the energy of the quark-antiquark pair. 

\fig{
Contour used to extract the quark-antiquark force from the Wilson loop.
The vertical direction indicates Euclidean time and the horizontal
direction indicates one of the spatial coordinates. This contour lives
in four dimensional Euclidean space. 
The parameters $\theta_{1,2}$ are for later reference.
}{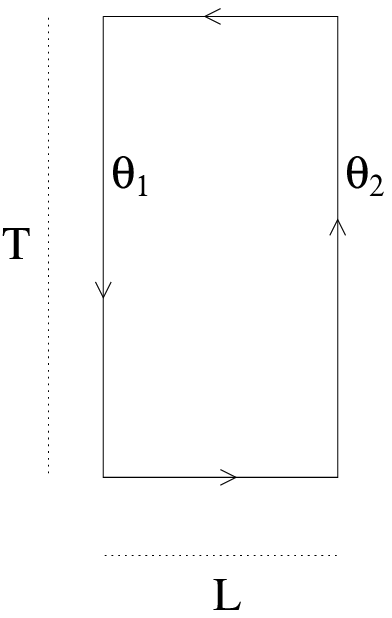}{2.5 truecm}
\figlabel\square

In order to perform this calculation for the  cases of interest
it will be necessary to introduce  massive quarks. 
To this effect consider breaking $U(N+1) \to  U(N) \times U(1)$ by 
giving some expectation value $\vec \Phi$ to a Higgs field. 
Then the massive W-bosons  have a mass proportional to
$|\vec \Phi|$ and  transform in the fundamental representation
of U(N).
So in the limit $|\vec \Phi| \to \infty$ they provide
the very massive quarks necessary to compute Wilson loops 
in the $U(N)$ theory. Notice that we are interested in 
physics for energy scales much lower than $|\vec \Phi|$ so that
that the $U(N)$ theory is effectively decoupled from the $U(1)$ theory.
Consider the equation of motion for the massive $W$ boson. Extracting
the leading time dependence as $W = e ^{-i |\Phi| t } \tilde W $ we
get an equation for $\tilde W$ which to first order in $1/|\vec \Phi|$
reads
\eqn\eofmo{
 ( \p_0 - i A_0  - i \theta^I X^I ) \tilde W =0
}
where 
we have defined $ \theta^I \equiv { \Phi^I \over |\vec \Phi| }$. 
Notice that $A_0$ and $X_I$ are matrices in the adjoint
of $U(N)$. 
This implies that if we consider this massive  $W$ boson describing a closed 
loop ${\cal C }$ its interaction with the $U(N)$ gauge field 
will lead to the insertion of the operator
\eqn\wilour{
W({\cal C }) = { 1 \over N} Tr P e^{ i \oint ds
[ A_\mu(\sigma) \dot \sigma^\mu + \theta^I(s)
X^I(\sigma)\sqrt{\dot \sigma^2 } ] }
}
The difference with \wilsonloop\ is the fact that we have an extra coupling
to $X^I$. 
The operator in \wilour\ is determined by the contour ${\cal C}$ 
(or $\sigma^\mu(s)$) as well
as a function $\vec \theta(s)$  mapping each point on the loop to a point
on the five-sphere. 
 We are interested in this operator because it
is the one that naturally arises when we consider the propagation of
a massive W-boson.
The appearance of $X^I$ might seem surprising at first sight, but
it is obvious when we remember that a string 
ending on a $p$-brane is not only a  source of electric field but 
it also carries ``scalar'' charge for  the fields $X^I$ since it 
is pulling the brane. In fact this coupling is crucial 
to understand the BPS bound for strings stretching between different 
branes \camabi . In the calculations below 
 $\theta(s)$ will be basically constant. 

\newsec{Relation to supergravity}

A natural proposal for the expectation value of 
the Wilson loop is 
\eqn\wiloursugra{
 \langle W({\cal C }) \rangle  \sim   e^{ -S}
}
where, in the large $gN$ approximation, $S$  is the
proper area of a fundamental string worldsheet which 
at  the boundary of $AdS$  describes   the loop ${\cal C}$ 
 and lies along    $\theta^I(s)$ on $S^5$.
See figure 2. In general we should consider the full 
partition function of string theory on  $AdS_5\times S^5$ 
  with the condition
that a string worldsheet is ending on the loop ${\cal C}$ and the points
$\vec \theta(s)$ on $S^5$ at the  boundary 
of $AdS$.
This is a natural proposal in terms of the 
identification proposed in \refs{\wittenhol,\gkp} for relating
gauge theory observables to  calculations on $AdS$.
However the 
right hand side in \wiloursugra\ contains also the
 contribution from the mass of the W-boson and it is therefore infinity.
Subtracting  this contribution we find 
a finite result for the Wilson loop operator
\eqn\wilfinal{
 \langle W({\cal C }) \rangle  \sim \lim_{\Phi \to \infty}   e^{ -(S_{\Phi} -
\ell \Phi )}
}
Where $\ell$ is the total length of the Wilson loop, measured with the
flat Minkowski metric appropriate to the gauge theory, and $\Phi$ is
the mass of the W-boson. The equation
\wilfinal\ is our final recipe for computing the Wilson loop.
This result is not ``zig-zag'' invariant, in the sense of \polyakov ,
since the operator \wilour\ is not invariant, as opposed to 
\wilsonloop .

\fig{
Proposal to 
calculate Wilson loop expectation values.
We should consider the partition function of string theory on 
$AdS_5\times S^5$ with a string worldsheet ending on the
contour ${\cal C}$ on the boundary of $AdS$. 
}{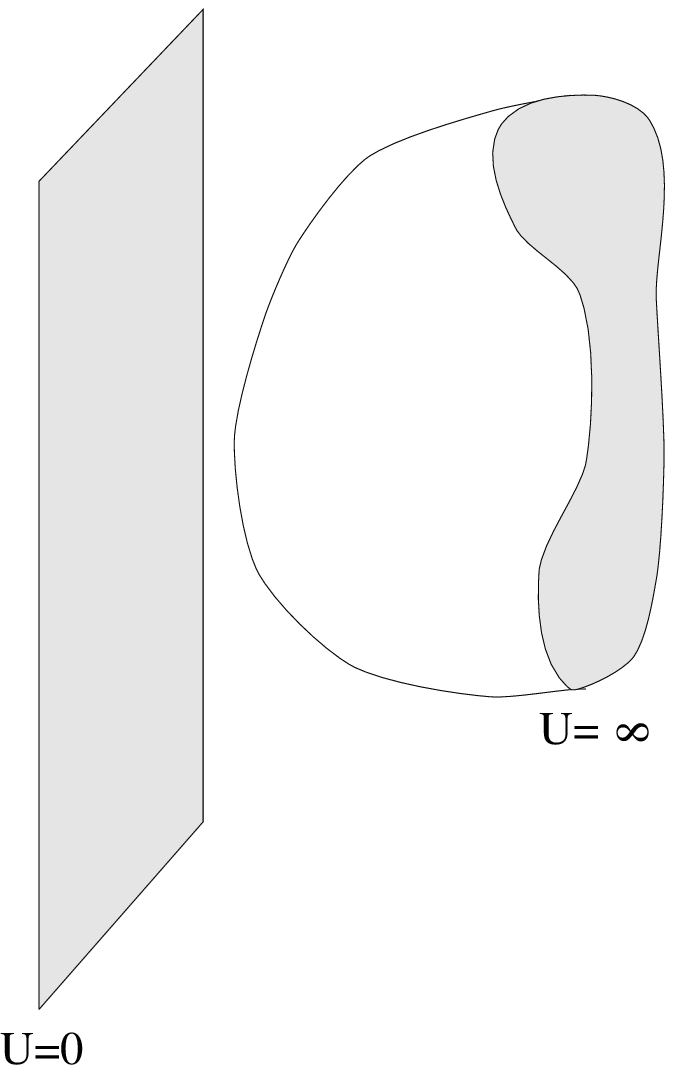}{2.7 truecm}
\figlabel\general

\newsec{ Quark anti-quark potential}

In this section we consider the calculation of a rectangular Wilson 
loop as  in  figure \square . We take the   angle 
$\theta^I(s)=\theta_0^I$ to be a constant. 
We consider the limit $T \to \infty$.
In this limit the problem becomes translational 
 invariant along the $\hat T$
direction. We put  the quark at  $x = -L/2$ and the
anti-quark at $x = L/2$. Here ``quark'' means an 
infinitely massive W-boson connecting the $N$ branes with
one brane which is far away in the direction $\vec \theta_0$.
The action for the string worldsheet is
\eqn\action{
S = { 1 \over 2 \pi \alpha'} \int d\tau d\sigma \sqrt{ \det G_{MN} 
\partial_\alpha X^M \partial_\beta  X^N }
}
where $G_{MN}$ is the Euclidean $AdS_5\times S^5$ metric
\eqn\metric{
ds^2 = \alpha' \left[
{ U^2 \over R^2} ( dt^2 + dx_i dx_i) + R^2 { dU^2 \over U^2 }
+ R^2 d\Omega_5^2
 \right]
}
where $R = (4 \pi g N)^{1/4}$ is the radius in string units and
$U = r/\alpha'$ has dimensions of energy.  
 Notice that
the factors of $\alpha'$ cancel out in \action , as they should.
Since we are interested in a  static configuration we take
$\tau =t, ~ \sigma = x$ so that the action  becomes
\eqn\act{
S = { T\over 2 \pi} \int dx \sqrt{ (\partial_x { U})^2+
U^4/R^4 } 
}
We need  to solve the Euler-Lagrange equations for this action.
Since the action does not depend on $x$ explicitly 
 the solution satisfies
\eqn\const{
{ U^4 \over  \sqrt{ (\partial_x { U})^2+
U^4/R^4 }}  = {\rm constant }
 }
Defining $U_0$ to be the minimum value of $U$, which by symmetry 
occurs at $x=0$, we find that the solution is\foot{All  integrals below 
can be calculated in terms of Elliptic of Beta  functions \gradsh .}
\eqn\sol{
 x = { R^2  \over U_0 } \int_1^{U/U_0} 
{ dy \over y^2 \sqrt{ y^4 - 1} }  
}
where $U_0$ is determined by the condition
\eqn\uzero{
{ L \over 2 } = { R^2  \over U_0}
\int_1^{\infty} 
{ dy \over y^2  \sqrt{ y^{4} - 1 } }  =  
{ R^2  \over U_0 } { \sqrt{2} \pi^{3/2} \over  \Gamma( 1/4)^2  }
}
The qualitative form of the solution is shown in figure 3.
Notice that the  string approaches the point $x =L/2$ quickly 
for large $U$
\eqn\approach{
{ L\over 2} -x \sim { 1 \over U^{3} }~,~~~~~U\gg U_0 ~ .
}
\fig{ (a) Initial  configuration corresponding to two W-bosons
before we turn on their coupling to the $U(N)$ gauge theory.
(b) Configuration after we consider the coupling to the $U(N)$ gauge theory.
This configuration minimizes the action. The quark-antiquark energy
is given by the difference of the total length of the strings in (a) and (b).
}{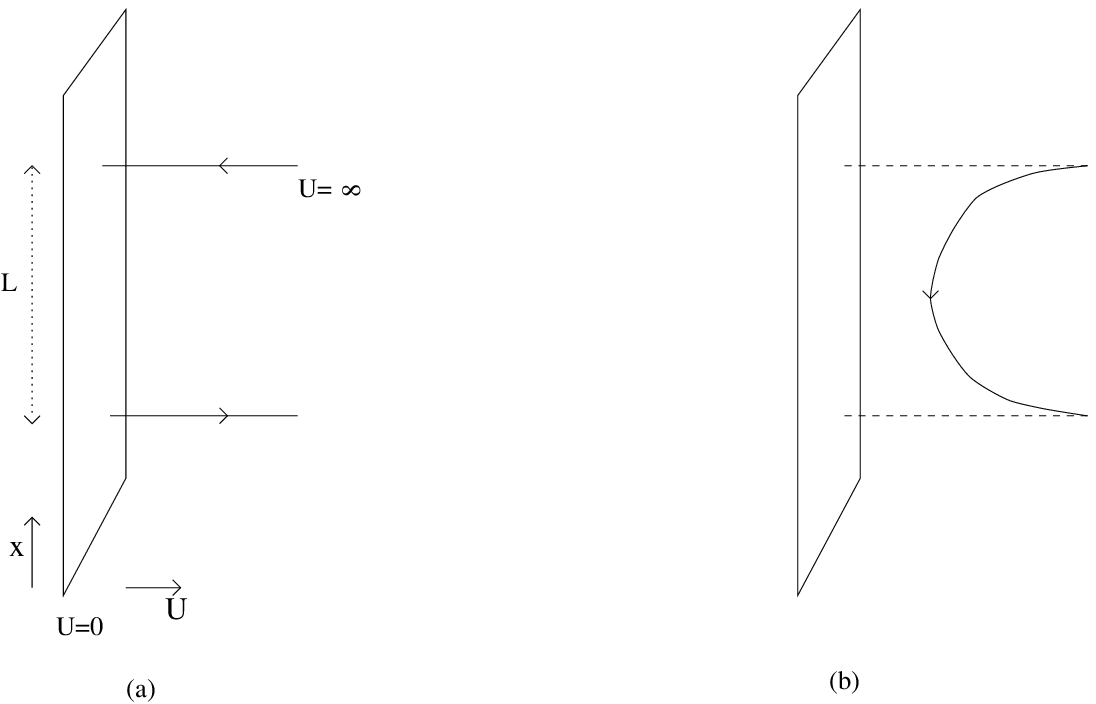}{10 truecm}
\figlabel\inout

Now we  compute the total energy of the configuration.
If we just plug in the solution \sol\ in \act , we find that
the answer is infinity. However as we said above
this infinity is simply due to the
fact that we are including the mass of the W-boson which 
corresponds to a string  stretching all the way to $U =\infty$. 
We can regularize the expression by integrating
 the energy only up to $U_{max}$.  Subtracting the regularized  mass
of the W-boson which is $U_{max}/(2\pi)$ we find a finite result\foot{
A convenient way to do the integral is to multiply the whole integral
by $y^\lambda$, calculate the two terms independently  as a function of 
$\lambda$ and then set $\lambda  =0$. }
\eqn\energy{\eqalign{
E =& { 2 U_0 \over 2 \pi }  \left[
\int_1^\infty dy \left( { y^2 \over \sqrt{  y^4 -1} }
 -1 \right)  -1 \right] \cr
E =&-  { 4 \pi^2 ( 2 g^2_{YM} N)^{1/2}  \over \Gamma({1 \over 4})^4 L} 
}}
We see that the energy goes as $1/L$, a fact  which is determined by
 conformal invariance. 
Notice that 
the energy goes as $(gN)^{1/2}$ as opposed to $gN$ which is the
perturbative result. 
This  indicates some screening of the charges. 
The above calculation makes sense for all distances $L$
when $gN$ is large independently of the value of $g$, this
suggest that one could define a magnetic Wilson loop operator
which for large $gN$ would be determined in terms of classical
D-string solutions with prescribed boundary conditions at infinity.
In the standard 't Hooft limit the interaction between Wilson loops
is  governed by $g $ which goes as $1/N$.

\subsec{Case of non-constant angle}

\fig{
Configuration of a $U(N+2)$ gauge theory Higgsed to $U(N)\times
U(1)_1\times U(1)_2$. This is a view in the transverse space. 
We also show two massive W-bosons they are characterized by the 
the angle of the Higgs expectation value of the $U(1)$ factor
that they are associated to. 
}{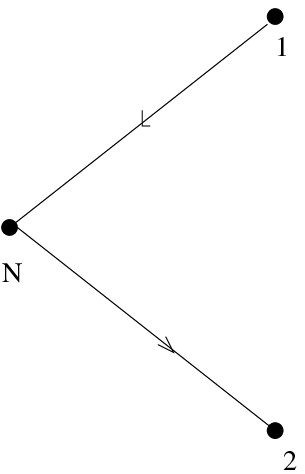}{2.5 truecm}
\figlabel\twow

Now we consider the case where the ``angle'' of the two quarks is
different. This arises when we break $U(N+2) \to U(N) \times U(1)_1
\times U(1)_2 $ by giving expectation values $\vec \Phi_1, \vec \Phi_2$
to the two $U(1)$ factors as indicated in figure \twow .
 Then the angles are $\vec \theta_i = \vec
\Phi_i/|\vec \Phi | $. So we consider a W-boson described by a string
going
between the $N$ branes and the brane associated to $U(1)_1$ and
a W-boson going between the brane associated to $U(1)_2$ and the $N$
branes as indicated in figure \twow . Notice that the orientation of the
string determines whether we have a quark (transforming in the
fundamental of $U(N)$)  or an anti-quark (transforming in the
antifundamental of $U(N)$).
The potential for this configuration can be calculated in terms of
the large $T$ limit of the expectation value of the Wilson loop shown 
in figure \square , with different values of $\vec \theta $ on each 
vertical line. 
 So we should consider a string worldsheet
which at $x = L/2$ goes to $U = \infty$ and to  the point  $\vec \theta_1$
of the five-sphere and at $x = -L/2$  goes to 
 $U = \infty$ and to the point  $\vec \theta_2 $ of the
five-sphere. The action for a time independent configuration is
\eqn\wsactang{
S ={ T\over 2 \pi }  \int dx \sqrt{ (\partial_x U)^2 + U^2 (
\partial_x
\vec \theta )^2 + U^4/R^4 } ~.
}
  From 
the symmetries of the problem we see that the string will lie along
a great circle of the sphere. So if we call $\theta$ the angle along this
great cicle we can choose $\theta_{1,2} = \pm \Delta \theta/2$. 
The problem then becomes symmetric around $x=0$. 
We can solve the Euler-Lagrange equations as above by using 
the fact that the Lagrangian \wsactang\ 
is independent of $x$ and $\theta$ so
that we have conserved quantities associated to ``energy'' and 
``angular momentum'' (interpreting 
$x$ as time).  
Solving these equations we find
\eqn\solution{\eqalign{
x = & { R^2 \over U_0 } \sqrt{1-l^2 }
 \int_1^{U/U_0} { dy \over y^2 \sqrt{ (y^2-1)(
y^2 + 1 - l^2  )} } ~ ,\cr
 \theta =&  l \int_1^{U/U_0} { dy \over \sqrt{ (y^2-1)(
y^2 + 1- l^2 ) } }~ ,
}}
and the parameters $U_0$, $l$ are determined by the 
conditions
\eqn\paramangl{\eqalign{
{L \over 2 } =& x(U = \infty)  ={ R^2 \over U_0 } \sqrt{1-l^2 } I_1(l) ~, \cr
{ \Delta \theta \over 2 } =&
\theta( U = \infty) = l I_2(l) ~,}}
where 
\eqn\integrals{\eqalign{  
I(l)_1 =& { 1 \over (1 - l^2) \sqrt{ 2 - l^2 } }\left[
(2 -l^2 )E\left({ \pi \over 2 }, \sqrt{1-l^2 \over 2 - l^2 }\right) - 
F\left( {\pi \over 2},  \sqrt{1-l^2 \over 2 - l^2 }\right) \right] ~,\cr 
I_2(l) =& { 1 \over \sqrt{2 - l^2 } } 
F\left( { \pi \over 2}, \sqrt{1-l^2 \over 2 - l^2 }\right) ~,
}}
with $F,E$ are elliptic integrals of the first and second kind. 
We can also calculate the energy of the system, substracting the mass of
the W-bosons and we find
\eqn\ener{\eqalign{
E =& 
{2  U_0 \over 2 \pi }  \left[ \int_1^\infty dy \left( 
{ y^2 \over \sqrt{ (y^2 -1) 
(y^2 +1 - l^2) } } -1 \right) -1  \right]
\cr
= &- { 2 \over \pi} { (2 g^2_{YM} N)^{1/2} \over L } (1-l^2)^{3/2} I_1^2(l)  
}}
Where $l$ is a function of the angle \paramangl .
It is interesting to notice that when $\Delta \theta \to \pi$ then 
$l \to 1$. Then  the solution looks like two straight strings going 
down to $U=0$ and the energy \ener\ goes to zero, as 
 expected
since this is a BPS configuration.

\fig{
Solution when the angles associated to the two W-bosons is different.
(a) shows the projection on the $x,U$ plane and 
(b) shows the projection on the $U,\theta$ plane, where $U$ now 
is the radial distance. 
}{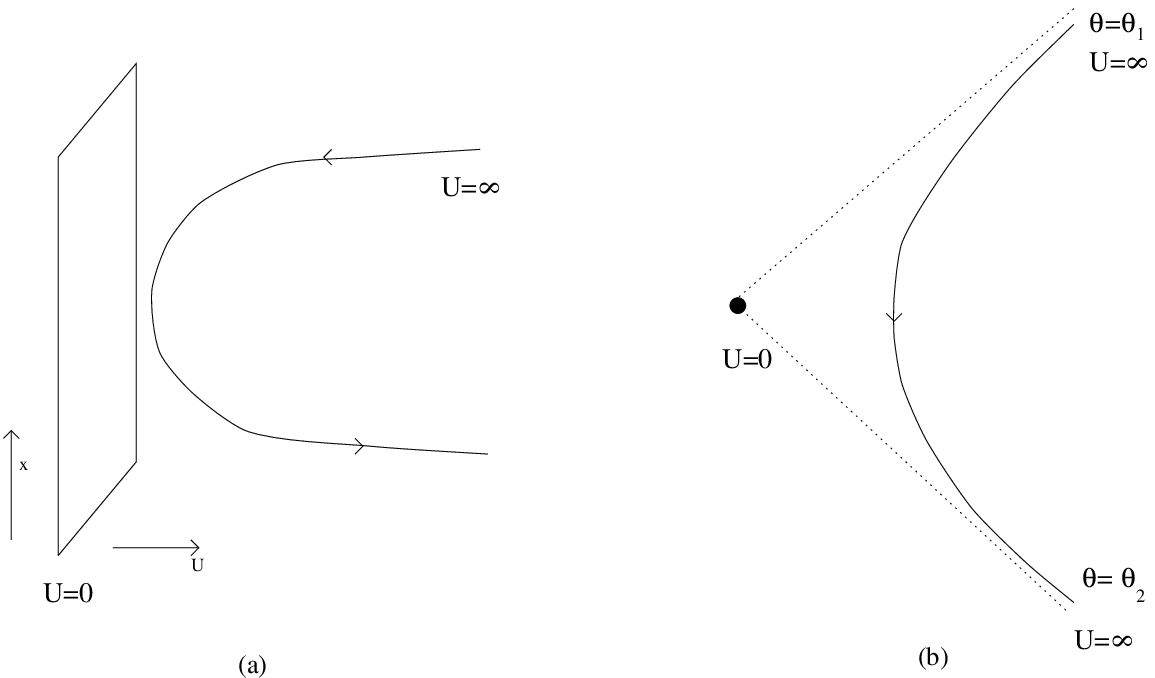}{10 truecm}
\figlabel\angle

\newsec{ M-theory membranes}

If we study the theory of coincident M-theory fivebranes,
the (0,2) conformal field theory in six dimensions \zerotwo , we are
led to consider M-theory on $AdS_7\times S^4$.
In this case one could define Wilson ``surface'' observables \ori .
Since we do not have an explicit formulation of the theory, 
we do not have a formula analogous to \wilour .
However we could define the Wilson ``surfaces'' as the phase factor 
associated to the propagation of a very heavy  string on 
branes (sustracting the part proportional to the free propagation of the
heavy string). In order to be more precise, let us suppose that
we start with $N+1$ branes and they we Higgs by separating one of the branes.
A membrane stretched between the $N$ five-branes at the origin and
the Higgsed five-brane  behaves  as  a string 
with  tension proportional
to the separation of the branes. We could consider this heavy  string
as probe for the unbroken conformal field theory associated with
the $N$ branes that are still together. The procedure is analogous
to what we saw above. The Wilson ``surface'' operator is defined
to be the extra phase factor associated with the interaction of
the heavy string with the $N$ fivebranes. 
This  Wilson ``area'' operator in the 
supergravity picture is defined by requiring that a membrane
ends at the boundary of $AdS_7\times S^4$ on the surface 
that defines the operator. Notice that we also have to specify a map from
the  surface
to  $S^4$ for the same reasons  described above for ${\cal N} =4$ 
 super-Yang-Mills.
 Again we  substract
the term corresponding to the free propagation of the heavy string to
obtain a finite result. For large $N$ we can trust the supergravity 
result. 

As an example, consider  a pair of parallel, infinite
strings corresponding to membranes 
ending on the fivebrane. Let us choose them with opposite 
orientation but  in the
same direction on $S^4$. This problem is translational invariant along
time and the direction of the strings. So the problem of determining
the minimal 3-surface reduces, as above, to finding  the
minimum of the action
\eqn\actionf{
S = { T L' \over (2\pi)^2 } \int dx \sqrt{ (\partial V )^2 + 
V^3/R^3 }
}
where now
 $R^3 = \pi N $ and $V=r/l_p^3$ has dimensions of (energy)$^2$.
 The strings have length $L'$ and  are separated by a distance $L$ in the
direction $\hat x$. We obtain the solution
\eqn\solmfive{
 x = { R^{3/2} \over V_0^{1/2} }  \int_1^{V/V_0} { dy \over
y^{3/2} \sqrt{y^3 -1} }~ 
}
where
\eqn\umf{
{L \over 2 }  = 
{ R^{3/2} \over V_0^{1/2} } { 2 \sqrt{\pi} \Gamma({2 \over 3}) 
\over \Gamma({1\over 6 }) }
}
If we calculate the energy we find
\eqn\energy{
{ E \over L'} = - { N \over L^2 } { 8  \sqrt{\pi} \Gamma({2\over 3})^3 
 \over 
\Gamma({ 1 \over 6})^3 }
}
The dependence on $L$ is the one expected from conformal invariance. 

\subsec{Wilson loops in non-conformal theories}

Consider 2+1 dimensional super-Yang-Mills with sixteen supercharges which
is the  theory  describing coincident  D2 branes.
We can define
the Wilson loop operator as in \wilour . 
Then we are lead to consider strings in the background of D2 branes.
The large $N$ limit of this  theory was considered in 
\imsy , where it was observed that the 
supergravity description is valid only in some region of the 
solution. Therefore the analysis of the Wilson loops will also 
be a bit more involved. 
We will find that we can calculate the Wilson loops from 
supergravity  only when the size of the loop is not too small.
This is just related  to the fact that for small distances 
we can trust the perturbative super-Yang-Mills theory. 
The physical result  is quite different when the Wilson loop is large. 
If we consider a string worldsheet, embeded in the $p$-brane
solutions studied in \imsy\ in a configuration appropriate for studing 
a the quark-antiquark forces we find that we have to minimize the
action 
\eqn\actionp{
S = { 1 \over 2 \pi } \int dx { \sqrt{ ( \p_x U)^2 + U^{5}/R^5 }}
}
where $R^5 = 6 \pi^2  g^2_{YM}N$.
We obtain solutions  very similar to \sol , which 
lead to the potential
\eqn\potq{
E =   -{ 2^{5/3}  \sqrt{\pi} \Gamma({4 \over 5} )^{5/3} 
\over 3^{1/3}  \Gamma({ 3 \over 10})^{5/3} } 
 { (g^2_{YM} N)^{1/3} \over L^{2/3} } = -{ \Gamma( {4 \over 5} ) U_0
\over \sqrt{\pi} \Gamma({ 3 \over 10}) }  
}
between quarks and antiquarks. $U_0$ is the minimum value of $U$. 
Now we perform the analysis of when we can trust \potq .
Let us first consider the large $U$ region. According to
 \imsy\ 
 we can trust supergravity for $U \ll g^2_{YM} N $. 
The solutions to \actionp\ consist of string worldsheets going
all the way to $U = \infty$. However the large $U$ behaviour of
the solution matches that of the infinitely massive W-boson.
So we will require the solution at  $U  \sim 
 g^2_{YM} N$ to be  very  similar to that of the W-boson, i.e. we require
 $x-L/2 \ll L $. This implies that
 $L \gg  1/(g^2_{YM} N) $. 
If the distance between the quarks was much smaller than the above bound
then we can apply perturbative Yang-Mills and we would
obtain a potential proportional to 
$V \sim   g^2_{YM} N \log (L g^2_{YM} N) $. We see that these
answers match up to a  numerical coefficient with \potq\ when both
calculations break down at $L \sim  1/(g^2_{YM} N) $.

Now we need to see if we can trust the behaviour of the
solution at small $U$, which corresponds 
to large distances.
%
%
 At small $U$ we expect that the
worldsheet of the string turns into an M-twobrane wrapped along
the eleventh direction. If $U_0 \gg  g^2_{YM}$ then we can trust the
above results \potq . 
If $U_0$ is smaller then we have to consider a more complicated situation
where we have to solve the equation of the M-twobrane in
the background corresponding to a periodic array of M-twobranes as
described in \imsy , this presumably could be done but we will not
attempt to do it here.

In summary, for the non-conformal theories one can also use 
the description of classical string worldsheets embeded
in some background supergravity solution 
 to calculate Wilson loops when the 
the size of the loop is large enough (otherwise we could
use  the perturbative 
description). 
One could consider other $p$-brane field theories as in \imsy\
and one would find various constraints on the size of the Wilson
loop for when one can trust the string worldsheet description.
Of course 
the total size is not the only issue, we also need that the contour
does not   wiggle too much. 

{\bf Acknowledgments}

I am grateful to the participants of the duality workshop at the 
 Institute for Theoretical Physics at  the University of California at
Santa Barbara for interesting discussions. 
I also thank N. Itzhaki for pointing out a typo in eqn. \potq . 

This work was supported in part by grants
DE-FG02-96ER40559 and NSF PHY94-07194.
 I thank the ITP at UCSB for hospitality.

\listrefs

\bye